\documentclass[11pt,namedreferences]{kluwer}

\usepackage{graphicx}

\def\'#1{\ifx#1i{\accent"13\i}\else{\accent"13#1}\fi}

\def\alamenos#1{$^{-#1}$}

\def\treceCO{{$^{13}$CO}}

\def\ltsima{$\; \buildrel < \over \sim \;$}    
\def\lesssim{\lower.5ex\hbox{\ltsima}}           
\def\gtsima{$\; \buildrel > \over \sim \;$}    
\def\gtrsim{\lower.5ex\hbox{\gtsima}}           

\def\apj{{ApJ}}
\def\apjs{{ApJS}}
\def\apjl{{ApJL}}

\def\aap{{AA}}
\def\mnras{{MNRAS}}
\def\nat{{Nature}}
\def\araa{{ARAA}}

\def\pasj{{PASJ}}
\def\pre{{Phys. Rev. E}}

\def\prom#1{\langle #1\rangle}

\begin{document}

\begin{opening}

\title{Turbulent Fragmentation and Star Formation}

\author{\surname{Javier Ballesteros-Paredes}}  

\institute{Instituto de Astronom\'ia, UNAM
\email{j.ballesteros@astrosmo.unam.mx}}

\runningauthor{J. Ballesteros-Paredes}

\runningtitle{Turbulent Fragmentation}

\begin{ao}\\
Javier Ballesteros-Paredes \\ Instituto de Astronom\'ia, UNAM \\
Ap. Postal 372, C.P. 58089 \\ Morelia Michoac\'an \\ M\'exico \\

\end{ao}

\begin{abstract}

We review the main results from recent numerical simulations of
turbulent fragmentation and star formation. Specifically, we discuss
the observed scaling relationships, the ``quiescent'' (subsonic)
nature of many star-forming cores, their energy balance, their
synthesized polarized dust emission, the ages of stars associated with
the molecular gas from which they have formed, the mass spectra of
clumps, and the density and column density probability distribution
function of the gas. We then give a critical discussion on recent
attempts to explain and/or predict the star formation efficiency and
the stellar initial mass function from the statistical nature of
turbulent fields. Finally, it appears that turbulent fragmentation
alone cannot account for the final stages of fragmentation: although
the turbulent velocity field is able to produce filaments, the spatial
distribution of cores in such filaments is better explained in terms
of gravitational fragmentation.

\end{abstract}


\end{opening}

\section{Introduction}\label{intro}

It is well known that stars form in dense cores within
molecular clouds, but the physical processes that control the
formation of low-mass stars within these cores are not well understood
yet. The ``standard'' scenario assumes that low-mass protostellar cores
are in quasi-static equilibrium supported against gravitational
collapse by a combination of magnetic and thermal pressures (see,
e.g., \opencite{Shu_etal87}). In this scenario, a core forms stars
once magnetic support is lost through the process known as ``ambipolar
diffusion'': neutral gas particles slowly drift through the ions, which
are held up 
by the magnetic field, allowing the core to eventually attain the ``critical''
mass-to-flux ratio, beyond which the gravitational energy exceeds the
magnetic energy and collapse sets in from the inside-out.

The more recent turbulent scenario, on the other hand, suggests that
cores are formed by compressible motions in the turbulent velocity
field of the cloud \cite{Sasao73, 
Elmegreen93,
Padoan95, BVS99, KHM00}. Those cores with an excess of gravitational energy
collapse rapidly to form stars, while the others with sufficiently
large internal or kinetic energies re-expand once the turbulent
compression subsides \cite{Taylor_etal96, VSB02}

Although the standard scenario has explained many observational
features of protostellar cores, it does not address the problem of how
the quasistatic cores can form and persist within a turbulent
environment. On the other hand, the turbulent scenario naturally
incorporates the process of core formation, and is increasingly
succesful in reproducing the observed core properties. In this paper,
we will review the main results of the turbulent scenario, the
observational facts that numerical works have successfully reproduced,
the simulations' predictions concerning the star formation efficiency
and the attempts to explain the initial mass function from statistical
considerations of a turbulent field. Finally, we discuss the final
stages of fragmentation, which appear to be gravitational rather than
turbulent.

\section{Turbulent Fragmentation}\label{intro:sec}

The process through which a chaotic velocity field produces a clumpy 
density structure in the gas within in a few dynamical timescales
is referred to as ``turbulent fragmentation''. 

It is generally accepted that small-scale turbulence can provide
support against global gravitational collapse (e.g.,
\opencite{Chandra_Fermi53}). In fact, if only small-scales modes
are considered, and if they are more or less isotropic, they provide
extra pressure. In this case, an ``effective'' Jeans length can be
defined in terms of the ``effective'' sound speed, $c^2_{\rm eff} =
c_s^2 + {1\over 3} u^3_{\rm rms}$, where $u_{\rm rms}$ is the
turbulent velocity dispersion, which in principle depends on the scale
\cite{Bonazzola_etal87}. Thus, if the scale of a system is smaller
than the modified Jeans length, it is said that turbulence inhibits
collapse. This is valid only if turbulence is isotropic, and
operates on small scales. In reality, interstellar turbulence is a
multiscale phenomenon, and its largest scales are in fact comparable
to the size of the system \cite{BVS99, Brunt03}, and contain most of
the kinetic energy. If this occurs, the large- and intermediate-scale
modes of turbulence can produce fragmentation, while the small-scale
modes can provide additional pressure. The structures produced by the
fragmentation (cores) can either proceed to collapse, or redisperse
and merge back into the ambient medium, depending on the balance
between the energies within them \cite{Sasao73, 
Elmegreen93, BVS99}. Thus, supersonic turbulence plays a
dual role, providing support for large structures while simultaneously
inducing their fragmentation into smaller-scale subunits. In what
follows, we review several predictions and comparisons with
observations from numerical simulations of turbulent fragmentation.



\section{Results}\label{predictions:sec}


\subparagraph{Turbulence decays fast.}
Molecular clouds (MC) are known to be turbulent, with motions that are
supersonic at scales $\ge 0.1$ pc \cite{Larson81}. Early concerns that
turbulence should be dissipated rapidly in shocks
\cite{Goldreich_Kwan74} 
were shortly afterwards argued to not be
applicable if turbulence consisted of MHD waves
\cite{Arons_Max75}. However, an important result found in numerical
simulations is that the original concerns were well founded, and that
strong MHD turbulence decays rapidly even in the presence of strong
magnetic fields (\opencite{Padoan_Nordlund99}; \opencite{ML_etal98};
\opencite{Stone_etal98}; \opencite{ML99};  \opencite{Avila_Vazquez01};
see, however, \opencite{Cho_etal02} for an alternative
viewpoint). Thus, it appears that MC turbulence must be continually
driven during the entire lifetime of the clouds (which, on the other
hand, probably does not exceed several Myr -- \opencite{BHV99}). The
driving mechanism, however, is probably not restricted to the stellar
activity internal to the clouds. Turbulence in clouds may be generated
by the very compressive process that forms the cloud, through
bending-mode instabilities aided by cooling \cite{Folini_Walder98,
Klein_Woods98, Koyama_Inutsuka01}, as proposed by \inlinecite{BVS99} and
\inlinecite{VBK03}.

\subparagraph{Energy balance.}
An evident result from numerical simulations is that the structures
formed in those models evolve dynamically because they are part of a
globally supersonic turbulent flow. Then, turbulence continuously
reshapes their structure, producing changes of mass, momentum and
energy in them \cite{BVS99}. Clouds in simulations are found to be
roughly in energy equipartition, but not in virial {\it equilibrium},
since the time derivatives of the moment of inertia and the surface
terms in their virial balance are at least as important as the volume
energy terms\footnote{This does not means that ``Virial Mass''
estimates are wrong. In fact, those estimations are based on
equipartition between gravitational and kinetic energies which, as
mentioned above, are roughly in equipartition. Instead, ``Virial
Mass'' estimations should be termed ``Equipartition Mass''
estimations.} (\opencite{BV95}; \opencite{BV97}; see
\opencite{Ballesteros02} for a review).  That the surface terms are
large implies the existence of strong fluxes of all physical
quantities (mass, energy and momentum) across cloud ``boundaries''
which, on the other hand, are rather arbitrary.

\subparagraph{Larson's Relations.}
Numerical simulations including turbulence have been helpful to
understand the origin of the observed scaling relationships, $\Delta v
\propto R^{1/2}$ and $\prom{\rho}  \propto R^{-1}$ \cite{Larson81}.
%
It is more or less accepted that the first relationship is a direct
consequence of the turbulent power spectrum of a field dominated by
shocks. However, it has been suggested that the second relationship
may be an artifact of the observations \cite{Kegel89, Scalo90}. In
fact, it has been shown from numerical simulations that the
relationship does not exist in real space (\opencite{VBR97};
\opencite{BM02}), but it appears in the observational
(position-position-velocity) space as a consequence of the
thresholding done in the process of defining clouds or clumps
\cite{BM02}.


\subparagraph{Fast cloud and star formation.}

If, as suggested by simulations, clouds are far from virial
equilibrium and supersonic turbulence decays fast, then probably cloud
formation can be faster than earlier estimates, which were $\sim$
20$-$30 Myr \cite{Blitz_Shu80}. This agrees with the suggestions by,
among others, \inlinecite{Sasao73},
\inlinecite{Hunter_etal86}, \inlinecite{Elmegreen93},
\inlinecite{BVS99}, that clouds are turbulent density fluctuations
within the global ISM flow.  

\inlinecite{BHV99} have shown that the global turbulence driven by star
formation events can produce locally gravitationally unstable clouds
with characteristics similar to those found in Taurus. They argue that
rapid assembly of molecular clouds by large-scale streams is a very
plausible mechanism to form clouds, in which, furthermore, stars older
than $\sim$ 5 Myr are not found\footnote{Note that
\citeauthor{Palla_Stahler00} (2000, \citeyear{Palla_Stahler02}) argue
that nearby MC have been forming stars in the last 10 Myr or more,
with a recent burst of star formation. However,
\inlinecite{Hartmann03} argues that their conclusions are skewed by a 
statistically small sample of stars with masses larger than 1$
M_\odot$, and by biases in their birthline age corrections. He notes
that the picture by \citeauthor{Palla_Stahler00},
\citeyear{Palla_Stahler02} (a) requires the last 1-2 Myr to be a
special epoch for most MCs; (b) implies that most MCs are forming
stars at extremely low rates, if any; and that (c) the apparently
oldest stars are systematically higher in mass, implying that for most
of a typical MC's star-forming history, the Initial Mass Function was
strongly skewed. Item (a) seems implausible, and items (b) and (c) are
contradicted by observations (see \opencite{Hartmann03} and references
therein).  A more plausible explanation of the observations, he
argues, is that the ``tail'' of older stars is really the result of
including older foreground stars, as well as problems with the
isochrone calibration in the higher mass stars.}. Indeed,
\inlinecite{HBB01} tabulate the ages of stars in 13 nearby
star-forming regions. For those regions with stars older than $\sim 5$
Myr, there is no molecular gas associated, suggesting that the time
scales for both cloud- and star-formation are short.

\subparagraph{Isolated and clustered modes of star formation.}

In numerical simulations, the structure of a turbulent molecular cloud
depends on the largest scale of the turbulent motions \cite{KHM00,
HMK01}. In particular, for turbulence driven at large scales,
``clustered'' collapse occurs predominantly in filaments in a more or
less efficient way, while for small-scale driven turbulence, a more
``isolated'' mode of star formation occurs (see \opencite{MK03} for a
review). However, a potentionally confusing point is worth noting
here. For the star formation community, a clustered region is one in
which many stars are formed in very compact regions. For instance,
while the Orion nebula has a star density of $\sim 10^4-10^5$ stars
per cubic pc, the Taurus Molecular cloud contains only about 200 T
Tauri stars. Thus, Orion is conventionally considered to belong to the
clustered mode, and Taurus to the isolated one. However, from the
point of view that Taurus is a coherent filamentary region, apparently
due to an external compression \cite{BHV99} that has formed stars
almost simultaneously, it would belong to the ``clustered'' mode as
defined by \inlinecite{KHM00}.


\subparagraph{Existence of Quiescent Cores.}
Observational evidence suggests that low-mass stars form from
molecular cloud cores whose column density profiles may resemble
Bonnor-Ebert \cite{Ebert55, Bonnor56} equilibrium spheres, and whose
velocity dispersions are transonic, or even subsonic \cite{Myers83,
Jijina_etal99, Caselli_etal02}. For this reason such cores are usually
termed ``quiescent''. In the standard scenario of star formation
\cite{Shu_etal87} quiescent structures are naturally explained as
consequences of the quasistatic contraction process. In the turbulent
scenario, however, protostellar cores are transient features generated
by the dynamical flow in the cloud. In a recent paper,
\inlinecite{BKV03} demonstrated that indeed, transient, dynamic cores
have an angle-averaged column density structure that often resembles
hydrostatic Bonnor-Ebert profiles, such as observed cores
\cite{Alves_etal01, Evans_etal01, Shirley_etal00,
Johnstone_etal00}. They argue that this kind of analysis in
observational work is not an unambiguous test of hydrostatic
equilibrium.

Furthermore, \inlinecite{KBVD03} investigate the velocity structure of
these cores. They exhibit sub- and transonic velocity dispersions and,
in self-gravitating cases are in approximate equipartition between the
kinetic and gravitational energies. They thus closely resemble the
observed population of so-called quiescent low-mass cores.  The fact
that dynamically evolving cores in highly supersonic turbulent flows
can have subsonic velocity dispersions is a direct consequence of a
turbulent energy spectrum that decreases with decreasing length
scale. Note that this also implies that the quiescent nature of the
cores is not necessarily indicative that they constitute the
dissipative scale of turbulence, as is sometimes suggested (e.g.,
\opencite{Goodman_etal98}).  The observed properties of quiescent
low-mass protostellar cores in molecular clouds thus do neither
directly imply the existence of strong magnetic fields for support
against collapse nor quasistatic evolution through ambipolar
diffusion.

\subparagraph{Polarized thermal dust emission.}
Other studies have focused on the polarized emission of dust from
numerical simulations. The main results are, first, that cores in the
magnetized simulations exhibit degrees of polarization between 1 and
10\%, regardless of whether the turbulence is sub- or super-Alfv\'enic
\cite{Padoan_etal01a}; second, that submillimeter polarization
maps of quiescent cores do not map the magnetic fields inside the
cores for visual extinctions larger than $A_V \sim 3$
\cite{Padoan_etal01a}; third, that the Chandrasekhar-Fermi method of
estimating the mean magnetic field in a turbulent medium
underestimates very weak magnetic fields by factors up to 150
\cite{Heitsch_etal01}; and fourth, that limited telescope resolution
leads to a systematic overestimation of the field strengths, as well
as to an impression of more regular, structureless magnetic fields
\cite{Heitsch_etal01}.

\subparagraph{Density and column density PDF of turbulent gas.}

Several numerical studies have investigated the statistical
distribution of the density field. The volume- (mass-) weighted
probability distribution function (PDF) of the density $f(\rho) d\rho$
is defined as the fractional volume (mass) occupied by gas in the
density range $(\rho,\rho+d\rho)$.

It has been shown that, for a polytropic gas (i.e., for a gas where
$P\propto \rho^\gamma$) the volume-weighted density-PDF depends on the
effective polytropic exponent $\gamma$ of the fluid. In particular,
for an isothermal flow, ($\gamma=1$), the PDF is lognormal \cite{VS94,
Padoan_etal97, PV98, Scalo_etal98, Ostriker_etal01, Li_etal03},
because the density fluctuations are the result of a stochastic
multiplicative process, and so the distribution of its logarithm is
Gaussian according to the Central Limit Theorem
\cite{PV98}. In the more general polytropic case ($\gamma\ne 1$),
\inlinecite{PV98} extended this result and showed that the PDF should
approach a power law at high (resp.\ low) densities for $\gamma < 1$
(resp.\ $\gamma >1$). Subsequently, other workers seem to have found
somewhat different numerical results. 
%
\inlinecite{Wada_Norman01} claim that a lognormal density PDF also
appears in their non-isothermal simulations of galactic disks, while
\inlinecite{Li_etal03} report that the shape of the PDF may depend on
the scale in which the turbulence is driven (see also
\opencite{Klessen00}. However, the discrepancy may be apparent. The
lognormal PDF of \inlinecite{Wada_Norman01} appears only at the
highest densities in their simulations, where they are nearly
isothermal, while the result of \inlinecite{Li_etal03} may be due to a
different choice of the turbulent driving, as already noted by
\inlinecite{PV98} and by \inlinecite{Klessen00}.

Finally, \inlinecite{PV03} have investigated the density PDF in the
magnetic isothermal case, together with the role of the magnetic
pressure, finding that the latter is only poorly correlated with the
density, because of the different scalings of the magnetic field
strength with the density fluctuations for the different types of
nonlinear MHD waves. This implies that, rather than providing a
restoring force, the magnetic pressure gradient acts more as a random
forcing, and therefore in general has little effect on the density
PDF, which in most cases retains its lognormal shape in the isothermal
case, as also found by \inlinecite{Ostriker_etal01}.

%
%

Since direct measurements of the density field in the ISM are not
available, it is important to understand the PDF of the column density
field. For that purpose, \inlinecite{Ostriker_etal01}
measured the PDF of the column density for numerical simulations,
finding 
%
that this distribution has a similar shape as that of the
underlying density field, but with smaller mean and width. 
However, \inlinecite{VS_Garcia01} argued that if the events are
separated by distances larger than a correlation length, 
then the distribution of the density
field should approach a Gaussian shape, although 
the convergence is very slow in the high density tail,
so that the column density PDF is not expected to exhibit a unique
functional shape, but to transit instead from a lognormal to a
Gaussian form, as the ratio of the column length to the decorrelation
length increases. This result, then, explains the nearly lognormal
distribution found by \inlinecite{Ostriker_etal01}.

\inlinecite{Blitz_Williams97} performed a comparison between column
density PDFs from models and observations 
by analyzing the PDF of the antenna temperature $T_A$ of
\treceCO\ for Taurus, assuming that
$T_A$ is proportional to the column density of the cloud. In their
Fig.~1, it is clear that the central region behaves closely to a
lognormal distribution. However, at larger $T_A$ the distribution
found for Taurus has an extended tail. This result may be due to the
fact that in Taurus self-gravity has produced high-density collapsing
cores. This is in fact apparent in the evolution of the volumetric
density PDF shown by \inlinecite{Klessen00} (see also Fig.~9 in
\opencite{Li_etal03}), it is clear that self-gravity allows for the
production of such tails in the volumetric density PDF. Thus, such
tails should appear also in the column-density PDF.

\subparagraph{Mass Spectra.}
The mass distribution of clouds and clumps in turbulent simulations
has been studied by several workers \cite{VBR97, Klessen01, BM02,
Gammie_etal03}. Their results generally give a mass spectrum of the
form $dN/dM \propto M^\alpha$, with $-2.5 \lesssim \alpha \lesssim
-1.5$. Since some of the results are discussed in the chapter by
Klessen (this volume), here we just comment that several studies have
recently tried to relate the mass spectrum of compact cores with the
slope of the mass distribution of the newborn stars (IMF). However,
this appears questionable because (a) the mass distribution of cores
changes with time as clumps merge together \cite{Klessen01}; (b) cores
generally produce a cluster of stars, not a single one, and so the
relation between the masses of cores and of individual stars is
unclear, and (c) many other processes are at play in determining a
star's final mass, such as competitive accretion (see, e.g., Bate,
this volume) and stellar feedback through winds and outflows. In other
words, a core with mass $M$ will not necessarily form, in general, a
star of the same mass, or even proportional to it.

\vskip -0.25cm
\subparagraph{The Star Formation Efficiency.}
Studies by \inlinecite{Padoan95} and \inlinecite{KHM00} have proposed
that turbulent fragmentation has an important role in determining the
Star Formation Efficiency (SFE). Moreover, the latter authors found
that the efficiency in numerical simulations (measured as the fraction
of mass in collapsed objects) increases as the energy injection scale
is increased.

As remarked by \inlinecite{Padoan95}, the compressible cascade regime
must end at a sonic scale, so that the turbulence within structures
smaller than this size is subsonic and thus cannot produce any
further subfragmentation. Thus, in this scenario, subsonic cores are
the natural end of the compressible part of the cascade in the
ISM\footnote{This does not necessarily imply that the dissipative (or
``inner'') scale of the turbulence has been reached, but only that at
this point, the cascade can turn into an incompressible
one. Similarly, it does not imply that there is a unique sonic
scale. A large scatter in the velocity dispersion of structures with a
given size is observed in simulations (e.g., \opencite{VBR97};
\opencite{VBK03}).}. 
Moreover, when turbulence becomes subsonic, it ceases to be a stronger
source of global support than thermal pressure, and the Jeans
criterion can be applied to determine whether a core is
gravitationally unstable and should proceed to collapse
\cite{Padoan95}, or else rebound and merge back into its environment 
\cite{Elmegreen93, Taylor_etal96, BVS99, VSB02}. This effectively sets
an upper limit to the SFE because if the cloud is globally supported
by its internal turbulence, only massive enough clumps among those
with sizes $\lambda$ smaller than the typical scale where the
turbulence becomes subsonic, $\lambda_s$ are susceptible of local
collapse \cite{Padoan95}. In this sense, star-forming cores constitute
the intersection of the set of subsonic regions with the high-mass,
super-Jeans tail of the mass distribution \cite{Padoan95, BV97,
Padoan_Nordlund02, VBK03}. In particular, \inlinecite{Padoan95} gave a
semiphenomenological theory which predicted the SF efficiency and mass
distribution of protostellar cores in turbulent MCs from the turbulent
flow parameters, using the Larson scaling relations as empirical
input. In that work, however, he used a questionable transformation
between density and Jeans mass (see \opencite{Scalo_etal98} for a
discussion).


In this scenario, since progressively smaller density peaks contain
progressively smaller fractions of the mass, one can expect the SFE to
decrease with decreasing sonic scale, the latter being a function of
the turbulent flow parameters \cite{VBK03}. Indeed, the latter authors
found that the star formation efficiency in numerical simulations
scales with $\lambda_s$ as $e^{-\lambda_0/\lambda_s}$, where
$\lambda_0$ is a reference scale, which for the set of simulations
fitted, has a value of $\sim$ 0.02~pc.

\subparagraph{The Initial Mass Function from turbulent
fragmentation.}\label{IMF:sec}
Interesting attempts to explain the Initial Mass Function (IMF) from
turbulent models have been performed by \inlinecite{Padoan_etal97} and
\inlinecite{Padoan_Nordlund02}. In particular, in the latter work,
%
%
%
%
%
the authors propose a model in which the power spectrum of the
turbulence is a power law, the cores are formed by shocks, and the
typical size of a dense core scales as the thickness of the postshock
gas; then the number of gravitationally unstable cores is given by

\begin{equation}
N(m) d\log m \propto m^{-3/(4-\beta)} \biggl( \int_0^m p(m_j) dm_j
\biggr) d\log m,
\label{padoan}
\end{equation}
where $\beta$ is the exponent of the energy power spectrum of the
turbulence, $E(k) \propto k^{-\beta}$, and the integral of $p(m_j)$ is
the fraction of cores of mass $m$ with gravitational energy larger
than the thermal energy. In particular, \cite{Padoan_Nordlund02} argue
that the value of $\beta=1.8$, which is close to the value found in
their numerical simulations, and is close to the average value
inferred from observations of the velocity dispersion-size
relationship, gives a mass dependence in eq.\ (\ref{padoan})
consistent with the Salpeter IMF.

\vskip -0.3cm
\subparagraph{Caveats in the derivation of the SFE and the
IMF.}\label{missing} 

%
%
%

As interesting and promising as it is, the model by
\inlinecite{Padoan_Nordlund02} is based on strong assumptions which we
briefly discuss here in order to asses their feasibility. First, the
number of cores as a function of scale, $N(L)$ (their eq. [17]) is
derived from scaling considerations as in \inlinecite{Elmegreen97}.
Thus, it is independent of estimates of core properties (eqs. [1]-[3],
[11] in \opencite{Padoan_Nordlund02}), while ideally, they should be
self-consistent. Second, the derivation of estimate of $N(L)$ is based
on assuming that a large-scale ($L_2$) simulation can be thought of as
made of $(L_2/L_1)^3$ space-filling simulations of scale $L_1$, thus
not taking into account intermittency of the density and velocity
gradient fields. Instead, it assumes that the small scales are
uniformly distributed in space without belonging to localized
larger-scale ones. Thus, it will be probably neccessary to do a
correction analogous to \inlinecite{Kolmogorov62} or $\beta$-model
intermittency corrections to original \inlinecite{Kolmogorov41} model.
This might alter the scaling. Third, their eq. (24), which gives the
number of collapsing cores, assumes that all collapsing cores are
subsonic and indivisible by turbulence. Specifically, if a core with
mass $m_1 < m_2$ is contained {\it within} a core of mass $m_2$, their
eq. (24) will predict two stars of masses $\propto m_1$ and $\propto
m2$, implying that the same mass used to form the first star, is also
used to form the second one. Moreover, recent numerical simulations of
the collapse of molecular cloud cores by \cite{Goodwin_etal03} have
found that the number and properties of objects formed in an
individual core depends in a chaotic way on the details of the initial
velocity field (assumed subsonic, with Mach $\sim$ 0.3), showing that
even subsonic cores fragment into a random number of stars. Fourth, in
their treatment, competitive accretion and stellar feedback is not
considered as agents that might change in a substantial way the
IMF. However, the interaction of protostellar cores and their
competition for mass growth from their surroundings are important
processes shaping the distribution of final star properties (see the
reviews by \inlinecite{MK03}, and \inlinecite{Larson03}). 

Finally, it is worth mentioning that in their treatment, both
\inlinecite{Padoan_Nordlund02} and \inlinecite{VBK03} consider that
the collapsing regions are those with subsonic turbulence and masses
larger than the Jeans mass. However, there are examples in our Galaxy
of compact ($R \le$~0.1~pc), massive ($M\sim 10^4 M_\odot$)
collapsing cores with velocity dispersions of the order of $\delta v 
\sim$ 4-12 km/sec (see, e.g., \opencite{Kurtz_etal00};
\opencite{Molinari_etal02}; \opencite{Faundez_etal03}). Although
it is likely that they might suffer from internal turbulent
fragmentation, it is not clear whether there will be one or several
(and how many) sites of collapse in them. In fact, $\sim 20$\%\ of the
objects observed by \inlinecite{Faundez_etal03} (e.g., IRAS
16272-4837) seem to be good candidates for isolated, massive star
formation regions.  Thus, these still need to be accommodated by the
emerging theory of turbulent star-formation.

\vskip -0.3cm
\subparagraph{Gravitational fragmentation on a filament produced by
large-scale streams}\label{Taurus:sec} 

One important question is whether there is again a transition from turbulent
fragmentation to gravitational fragmentation at some small scale, and
if so, what is that scale.

In the particular case of Taurus, once turbulence has induced
the formation of the filamentary structure of the gas 
\cite{BHV99}, gravitational fragmentation in the
filaments should be considered as a mechanism for forming cores and
stars. In fact, following \inlinecite{Larson85},
\inlinecite{Hartmann02} has shown that the Jeans length and the
free-fall timescale  for a filament, are given by 

\begin{subequation}[arabic]
\begin{eqnarray}
\lambda_J  = & 1.5 \ T_{10}\ A_V^{-1} \ {\rm pc,} \label{lambda_lee:eq} \\
\tau   \sim  & \ 3.7 \ T^{1/2}_{10}\ A_V^{-1} \ {\rm Myr.}
	\label{timescale_filament:eq}
\end{eqnarray}
\end{subequation}
%
where $T_{10}$ is the temperature in units of 10~K, and $A_V$ is the
visual extinction through the center of the filament. 
%
By using a mean visual extinction of starless cores $A_V\sim 5$
\cite{Onishi_etal98}, eq. (\ref{lambda_lee:eq}) gives a characteristic
Jeans length of $\lambda_J\sim 0.3$~pc, and collapse should occur in
about 0.74~Myr. \inlinecite{Hartmann02} argues that, since the number of
young stellar objects per length scale in Taurus is about $3-4$
pc\alamenos 1, then self-gravity is working and producing the
cores and stars.

\begin{figure}[H]
\centerline{\includegraphics[width=27pc]{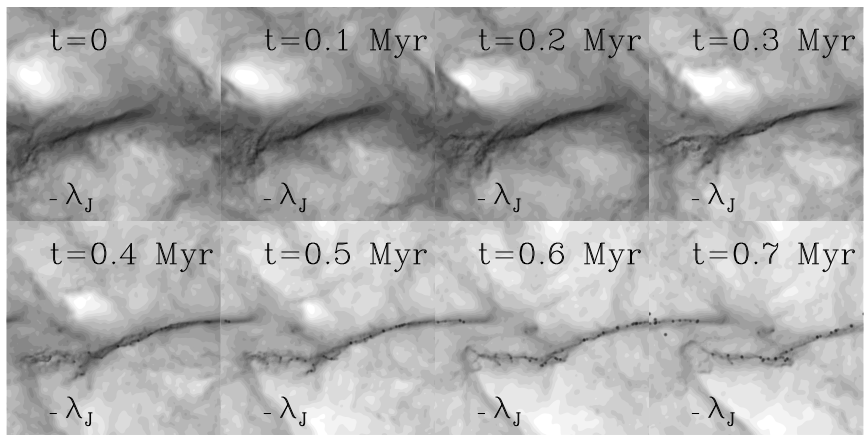}
\caption{Evolution of the column density of an SPH simulation
(courtesy of R. Klessen). The
filament in the first frame (before self-gravity is turned-on) shows
that turbulence is responsible for forming this kind of structures. The
small bar in the bottom-left of each frame denotes the Jeans length
(eq. [\protect\ref{lambda_lee:eq}]) at this time. At later times,
self-gravity is turned on and the filament suffers gravitational
fragmentation on a free-fall timescale
(eq. [\ref{timescale_filament:eq}]). The size of each box is 6~pc.}
\label{filament:fig}
}
\end{figure}

In order to test this idea in the simulations, we analyze an SPH
calculation performed by R. Klessen (see his contribution in this
volume for more details). This simulation has been performed without
gravity until a particular, well defined elongated structure is
formed. At this time, self-gravity is turned on. The results are in
good agreement with \inlinecite{Hartmann02}. In
Fig.~\ref{filament:fig}, we show eight column density frames of the
simulation. The first frame shows the structure just before
self-gravity is turned-on. We see that the filament forms cores in a
fraction of a Myr. The timestep between frames is 0.1~Myr. The mean
surface density for the filament is 0.033~gr~cm\alamenos 2,
corresponding to a visual extinction of $\sim$~7.5. Using
eqs. (\ref{lambda_lee:eq}) and (\ref{timescale_filament:eq}), this
value gives a Jeans length of $\lambda_J \sim 0.2$~pc, and a
collapsing timescale of $\tau \sim 0.5$~Myr. Note from
Fig.~\ref{filament:fig} that the first cores appear roughly at $\tau
\sim 0.3$~Myr, although the final structure of collapsed objects is
clearly defined at $t=0.5$~Myr. Moreover, the typical separation
between the cores corresponds roughly to the Jeans length, which is
indicated by a small bar in the figures.


We then conclude that even though the filament is not in hydrostatic
equilibrium, gravitational fragmentation may produce the structures
seen either in the simulations, and in actual clouds. This means that
turbulence causes first the production of the filament, and then
self-gravity takes over to form the collapsed objects. However, since
(a) the filament is not a perfect cylinder, (b) the collapsed objects
are not perfectly equally spaced, and (c) there is a range in
timesteps in which the cores are formed (between $t \sim$~0.3 and
0.6~Myr), it becomes clear that the final properties of the star
forming regions still depend on the initial conditions induced by the
global turbulence. In a similar way, the final mass of the stars may
still depend on the accretion history from the surrounding cloud
material (see chapter by Klessen, this volume).

\vskip -0.5 cm

\section{Conclusions}

This review shows that numerical simulations of turbulent
fragmentation have been quite successful in reproducing a number of
observational properties of molecular cloud cores, while
simultaneously making a number of predictions about the behavior of
turbulence in molecular clouds and their cores, as well as providing
an explanation for another set of observed features. Although much
work remains to be done, and a full theory is still elusive, the
progress made is highly encouraging, and makes the dynamical
picture of star formation a plausible alternative to the
ambipolar-diffusion-mediated magnetostatic-equilibrium standard model.

\vskip -1 cm
\acknowledgements
I want to thank Enrique V\'azquez-Semadeni and Ralf Klessen for
fruitful discussions and careful reading of this manuscript; the
Scientific Organizing Committee for their invitation; Paolo Padoan for
interesting (and still ongoing discussions) on the derivation of the
IMF from the mass distribution of clumps formed by the turbulent
field; and support from Conacyt's grant I39318-E.

\end{document}